\documentclass[preprint]{aastex}

\shortauthors{Narlikar et al.}
\shorttitle{MBR inhomogeneities in QSSC}

\input epsf

\begin{document}


\title{ Inhomogeneities in the Microwave Background Radiation
interpreted within the framework of the Quasi-Steady State Cosmology}

\author{J.V. Narlikar\altaffilmark{1}, R.G. Vishwakarma\altaffilmark{1},
Amir Hajian\altaffilmark{1,2}, Tarun Souradeep\altaffilmark{1},
G. Burbidge\altaffilmark{3} and F.~Hoyle\altaffilmark{4} }

\altaffiltext{1} { Inter-University Centre for Astronomy and
  Astrophysics, Post Bag 4, Ganeshkhind, Pune 411 007, INDIA.}
\altaffiltext{2} { Institute for Advanced Studies in Basic Sciences,
  P.O. Box 45195-159, Zanjan 2354, IRAN} 
\altaffiltext{3} {Center for Astrophysics and Space Sciences 0424,
  University of California, San Diego,
  CA 92093-0424, USA}
\altaffiltext{4} {Deceased}



\begin{abstract}
  
  We calculate the expected angular power spectrum of the temperature
  fluctuations in the microwave background radiation (MBR) generated
  in the quasi-steady state cosmology (QSSC).  The paper begins with a
  brief description of how the background is produced and thermalized
  in the QSSC.  We then discuss within the framework of a simple
  model, the likely sources of fluctuations in the background due to
  astrophysical and cosmological causes.
  Power spectrum peaks at $l \approx
  6-10$, $180-220$ and $600-900$ are shown to be related in this cosmology
  respectively to curvature effects at the last minimum of the scale factor,
  clusters and groups of galaxies. The effect of clusters is
  shown to be related to their distribution in space as indicated by a toy
  model of structure formation in the QSSC.
  We derive and parameterize
  the angular power spectrum using six parameters related to the
  sources of temperature fluctuations at three characteristic scales.
  We are able to obtain a satisfactory fit to the observational band power
  estimates of MBR temperature fluctuation spectrum.  Moreover, the
  values of `best fit' parameters are consistent with the range of
  expected values.

 
\end{abstract}



\section{INTRODUCTION}

The quasi-steady state cosmology (QSSC) has been proposed by Hoyle,
Burbidge and Narlikar (1993, 1994$a$, $b$, 1995) as an alternative to
the standard hot big bang model.  This cosmology does away with the
initial singularity, and does not have any cosmic epochs when the
universe was very hot. The synthesis of light nuclei and the origin of
the microwave background radiation (MBR) have therefore been explained
by different physical processes than those invoked in the hot big
bang. (cf. Hoyle, Burbidge and Narlikar, 2000; Burbidge and Hoyle,
1998).

We concentrate here on the MBR and our purpose is to explain its
observed fluctuations in terms of the large scale features of the
QSSC.  First we begin with a brief description of how it is generated
in this cosmology.  A typical QSSC scale factor for a flat
Robertson-Walker universe is given by

\begin{equation}
S(t) = e^{t/P}  [{1~+~\eta {\rm ~cos} (2\pi \tau/Q)}],
\end{equation}

\noindent 
where the time scales $P~\approx 10^{3}$ Gyr $\gg Q~\approx 40-50 $
Gyr are considerably greater than the Hubble time scale of 10-15 Gyr
of standard cosmology.  The function $\tau(t)$ is very nearly like
$t$, with significantly different behaviour for short duration near
the minima of the function $S(t)$. The parameter $\eta$ has modulus
less than unity thus preventing the scale factor from reaching zero.
Typically, $\eta \sim 0.8-0.9$.
Hence there is no spacetime singularity, nor a violation of the law of
conservation of matter and energy, as happens at the big bang epoch in
the standard model.  This is because, matter in the universe is
created through minibangs, through explosive processes in the nuclei
of existing galaxies that produce matter and an equal quantity of a
negative energy scalar field $C$.  Such processes take place whenever
the energy of the $C$-field quantum rises above the threshold energy $m_{\rm
  P} c^2$, the restmass energy of the Planck particle which is
typically created.  For details of the basic physics see Hoyle et al
(1995) and for models of the kind (1) see Sachs et al (1997).

While the overall energy level of the $C$-field is below this
threshold, it can rise in the neighbourhood of highly collapsed
massive objects (objects close to becoming black holes), and these
then become the sites of creation.  Also since the overall energy
density of the $C-$field goes as $S^{-4}$, the creation is facilitated
near the epochs of smallest $S$.  It is close to these relatively
denser epochs that the creation activity is at its peak, switching
itself off as the universe continues to expand in the cycle.  Hence
from one cycle to another the density of matter would have fallen by
the factor exp$(-3Q/P)$, but for the compensating creation of matter
at the beginning of a new cycle at the minimum of $S$.

Thus in the simplest model, although the universe is in a long term
steady state, it oscillates over shorter time scales, with {\it each
  cycle physically the same as the preceding one}.  Each cycle has the
new matter taking part in the process of formation of stars and
galaxies which evolve exactly as in the previous cycle.  What happens
to the radiation of stars in this process?  Assuming that the
radiation density falls like $S^{-4}$ in each cycle, it would be
depleted by the factor exp$(-4Q/P)$ from one oscillatory minimum to
the next.  It is this gap that is made up by the starlight generated
during the cycle, since the universe is in a steady state from one
cycle to the next.  Given $P$ and $Q$, and the starlight energy
generated per cycle, we can estimate the radiation background at any
stage of a cycle.  We have shown elsewhere (Hoyle, Burbidge and
Narlikar 1994a, 2000) that it is this starlight thermalized by dust
grains which is responsible for the microwave background (MBR) in the
QSSC.  Further, the estimate of starlight generated
during the typical cycle allows one to show that after thermalization
the temperature at the present epoch should be very close to $2.7$ K
(Burbidge and Hoyle 1998; Hoyle et al 2000).
 
In the following sections we first review the spectrum and homogeneity
of the MBR in the QSSC, and then describe the likely causes of
inhomogeneity of the MBR. We finally compare the results with recent
observations in which the fluctuations of the MBR at different angular
scales have been measured.

\section{THE SPECTRUM AND TEMPERATURE OF THE MBR}

In this model it is hydrogen burning in stars in galaxies that
generates the energy of the MBR, and the energy distribution of
stellar radiation over different wavelengths from a typical galaxy
varies through the cycle in the following way.  At the minimum $S$
epoch, new gaseous material is acquired, and new stars are formed,
with the energy coming out mostly in the blue to ultraviolet, from
stars of masses greater than $M_{\odot}$.  However, as the cycle
proceeds towards the maximum $S$ phase after the elapse of times of
the order $\approx 25$ Gyr, the massive stars burn out, leaving only
the low mass stars still shining.  The typical stars thus left are the
dwarfs of types $K$ to $M$, with the consequence that the radiation is
then mainly in the red and the infrared.  However, as the cycle
proceeds towards the next minimum, these wavelengths are shortened
since the universe contracts by a considerable factor.

The total energy density of starlight generated in a given cycle can
be estimated from the present observations of the starlight background
and by extrapolating the star burning activity over the full cycle.
Suppose that this energy density is $\epsilon$.  If the background
radiation energy density $u$ at the start of a cycle (when the scale
factor was at its minimum value $S_m$) were $u_m$, then in the absence
of any new addition to it, following the $u \propto S^{-4}$ law, this
would drop to $u_{m} {\rm exp}(- 4Q/P)$ at the end of the cycle.  To
maintain a steady state from one cycle to next therefore the shortfall
in $u_m$ is made up by $\epsilon$.  Hence

\begin{equation}
\epsilon = u_m\{1 - e^{-4Q/P}\}.
\end{equation}

This fixes the value of $u_m$ in terms of starlight energy input.  If
we know the maximum redshift $z_m$ in the present cycle, we find that
the present energy density of the MBR is

\begin{equation}
u_{o} = u_{m}~(1+z_{m})^{-4}.
\end{equation}

The details of $\epsilon$, $z_m$, P and Q were discussed in the papers
cited above (see, for example, Hoyle et al 2000).  A present day MBR
temperature of $\approx 2.7 K$ arises from such considerations.
However, to talk of a `temperature' one must first show that the MBR
has reached thermal equilibrium.  We next describe briefly what must
occur if this result is to be achieved.

To fix ideas, we will consider the typical oscillatory cycle from one
minimum of scale factor to next. Let

\begin{equation}
x = S(t)/S(t_{min})
\end{equation}

\noindent 
where, in going from one minimum to the next we have ignored the
exponential term. It is clear from this time dependence, that in going
from the maximum to the minimum, the scale factor drops by a factor
$(1~+~\eta)/(1~-~\eta) \approx 12$ for $\eta \approx 0.85$.  We shall
assume as an additional parameter that the present epoch is
characterized by $x~=~6$.

Thus, infrared wavelengths of 1000 nm at $x=6$ are shortened to
ultraviolet wavelengths of only $\approx 160$ nm at the $x=1$ (minimum
$S$) epoch.  However, by this stage, the intergalactic density is
increased by a factor $6^{3}$ as the minimum is approached and the
issue of absorption becomes important. For it is through frequent
absorptions and re-emissions that the radiation background gets
thermalized.

We have shown that metallic particles in the form of graphite whiskers
as well as iron whiskers play a crucial role in the absorption
process. As Narlikar et al (1997) have discussed, a plausible case
based on laboratory experiments and astrophysical evidence, can be
made for the condensation of metallic whiskers from the hot ejecta of
supernovae, which are blown out into intergalactic space by the shock
waves generated at the time of explosion.  The extinction properties
of such whiskers, typically of length $\approx 1$mm and diameter
$0.01\mu$m differ considerably from those of normal spherical dust.
In particular the iron whiskers at cryogenic temperatures are a
dominant source of opacity in the microwave region, while carbon
whiskers are more effective at the shorter UV wavelengths.

Typically a galaxy belongs to a cluster and we expect these whisker grains to
fill up the intergalactic space within the cluster. The starlight from
a galaxy in the cluster will therefore pass through such a dust as it
emerges out of the cluster. We expect therefore absorption and re-radiation
of starlight outside galaxies and also, in a larger angular scale, outside
clusters. As we shall see later, the production of microwaves in this
fashion will go on in each cycle, and the process of frequent absorption
and re-radiation by whiskers will eventually generate a uniform background,
{\it except for the contribution from the latest generation of clusters}.
These will stand out as inhomogeneities on the overall uniform background.

The value of absorption coefficient $Q_{{\rm abs}}$ for graphite
whiskers is essentially constant for all wavelengths longer than
$\approx 1 \mu$m, extending even to the long radio wavelengths, being
equivalent to an absorption coefficient of $\sim 10^5$cm$^2$g$^{-1}$.
However, it is three times this value for the UV radiation.  Now an
intergalactic density of $\approx 10^{-34}$ g cm$^{-3}$ at $x=6$ would
rise to $\approx 2\times 10^{-32}$g cm$^{-3}$, at the minimum ($x=1$)
epoch.  Over a cosmological distance of $10^{27}$cm at this stage, the
optical depth would be $\approx 6$ for UV, and $\approx 1$ for
wavelengths longer than 1 $\mu$m.

It is significant that this type of dust plays a role in the observed
redshift apparent magnitude relation of Type Ia supernovae.  As shown
by Banerjee et al (2000), an intergalactic dust density of the above
order gives a very good fit to the observations (Perlmutter et al
1999, Riess et al 1998).  Since the results are very sensitively
dependent on dust density, it cannot be fortuitous that the density
for `best fit' in the supernova case is of the right order to explain
thermalization of the MBR.  Further, recent studies of the high
redshift supernova SN 1997$ff$ ($z \cong 1.7$) show the model to be
consistent with observations within the error - budget currently
applicable (Vishwakarma 2002, Narlikar et al 2002).

The great bulk of the optical radiation that becomes subject to
thermalization in the contracting phase of the oscillation will have
traveled a distance of the order of 10 Gpc, or even more in the case
of microwaves. The radiation incident on a carbon whisker will have
been in transit since the maximum $S$ epoch of the previous cycle, and
it includes all the microwave radiation existing {\it before} the
present cycle, as well as the starlight generated by the galaxies in
the current cycle.  The result is that all this radiation is well
thermalized and uniform in energy density. However, the carbon
whiskers themselves will be lumpily distributed, on the scale of
clusters of galaxies.  So, as the minimum of $S$ is approached at the
end of the last cycle, the conversion of starlight to microwaves will
have been lumpy.  As the starlight is progressively absorbed, the
typical grain temperature $T_g$ first rises above the MBR temperature
$T_{{\rm MBR}}$ before dropping back to it as the grain radiates.

The effect of grains being lumpily distributed
on the MBR is to cause the slight rise followed
by the fall back to the MBR temperature to be lumpy too. The radiation
background itself, however, does not have this lumpiness: because the
total assembly of grains has a negligible heat content and in thermal
equilibrium, each grain emits as much heat as it receives.  Since the
emissivity of the particles does not depend on wavelength, each emits
a Planckian spectrum corresponding to its temperature $T_g$.
Eventually when all particles come down to the temperature $T_{{\rm
    MBR}}$, however, further absorption and reemission bring about a
strict Planckian distribution at this temperature.  Although as was
suggested by Narlikar, Wickramasinghe and Edmunds (1975), the
thermalization can be achieved by the carbon whiskers, the presence of
a small quantity of iron whiskers helps the process further.  The
absorption and re-radiation by the whiskers at the oscillatory
minimum, thus generate a mixing of radiation from distances as far as
$\approx 10^{29}$ cm, at the relatively low intergalactic particle
densities prevalent there, so in effect permitting radiation to travel
freely.

\section{THE ORIGIN OF MBR INHOMOGENEITIES}

This picture suggests that the overall background will be very smooth.
However, there will be some tiny fluctuations of intensity stamped on
it arising from certain intrinsic inhomogeneities of the process as
well as the cosmological model.  We consider them in decreasing order
of angular scale.

For this purpose we propose a simple model in which the sources of
inhomogeneities can be traced to the epoch of last minimum of
oscillation ($x= 1$ in our present case).  This is the maximum density
epoch when the extinction by dust will be most effective.  The
characteristic geometrical size of the model is determined by the
spacetime curvature $R$ and will be of the order $R^{-1/2}$.  Any
large scale inhomogeneity of the MBR would arise over this
characteristic cosmological size. This size features in standard
cosmology also and is referred to as the `horizon size' or the
`Hubble radius' ({\it see} Weinberg 1972).
For smaller sizes we need to take note of
inhomogeneity on the scale of rich clusters of galaxies, since they
represent local concentrations of starlight not yet fully thermalized
and redistributed.
On still smaller scales we expect to see inhomogeneities on the scale
of groups of galaxies or even individual galaxies.  (There should
likewise be some inhomogeneity on the scale of superclusters, but it
is expected to produce a weaker effect and we will ignore it in the
present model.)

It is clear from the above that the issue of inhomogeneities of the
MBR in the QSSC is linked with the way large scale structures are
formed in this cosmology.  A structure formation scenario in the QSSC
has not been developed to the level of sophistication that it has
acquired in standard cosmology.  A toy model by Nayeri et al (1999)
has shown, however, that the structure formation process as in the QSSC
is very different from that in the standard cosmology, being
essentially driven by the minicreation events with gravitation playing
a secondary role.  Thus existence of galaxies and clusters at high
redshifts ($z \sim 5$) is taken for granted in the QSSC picture of
structure formation.  The toy model of Nayeri et al shows how
clustering at various scales develops from an initial random
distribution, through creation events and expansion.  In a `steady
state' situation the model has to demonstrate how the physical
conditions in one QSSC cycle are repeated in the next.
We will elaborate on this idea in the following section.


To give quantitative estimates of the above effects, we will
choose the same cosmological parameters that we have used to
explain the redshift magnitude relation based on Type Ia supernovae
({\it see} Banerjee, et al, 2000).  They are :

\begin{equation}
H_0 =65 {\rm ~km ~ s}^{-1}~{\rm Mpc}^{-1}, ~~~ z_{{\rm max}} = 5,
~~~\Lambda_0~=~-0.358.
\end{equation}

The parameters $P$ and $Q$ do not enter explicitly in the calculation
but we may typically take them to be 1000 Gyr and 50 Gyr,
respectively.

The largest angular scale will arise from the size of spacetime
curvature.  At the present epoch this will be of the order of
$(c/H_{o})^{-2}$.  This curvature corresponds to a linear size
$c/H_{o}$.  As we go towards the minimum scale factor epoch, this
scale decreases.  The angle subtended by the scale at us from the
minimum scale epoch can be worked out.  For the parameters $P \approx
1000$ Gyr, $Q \approx 50$ Gyr, $z_{{\rm max}}=5-6$, we get an
angular size of $\approx 10^\circ$.  We therefore expect this scale to
show up in the regime of relatively large angular scale fluctuations of the
intensity of MBR.  The first discovery of inhomogeneity of MBR by COBE
(Smoot et al 1992) was of this order.

\medskip

 To assess the magnitude of fluctuations, we reproduce the arguments
 of Hoyle, at al (1994a). Consider the whisker grain distribution at the
 epoch of last minimum of $S(t)$. The whiskers may be distributed
 non-uniformly to begin with, with some regions having larger whisker
 densities than others. Until the deviations from uniformity of radiation
 background become too small, they are able to push the grains down the
 temperature gradients to restore uniformity. Consider a fluctuation by
 a small factor $y$ in the average energy density $\sim  5 \times 10^{-10}$
 erg cm$^{-3}$ of the microwaves at this stage. A fluctuation of this
 order is able to {\it move}  grains of density $\rho_g$ and velocity $V$
 provided $1/2 \rho_g V^2 \cong 5\times 10^{-10}y$. With
 $\rho_g\sim 2\times 10^{-33}$ g cm$^{-3}$ and $V\approx 0.1 c$ (an adequate
 speed to fill a region of size $\sim  10^{26}$ cm in the available
 cosmological times of the order $10^{17} s$), we get
 $y \approx 2\times10^{-5}$, corresponding to $\Delta T/T\approx 5\times 10^{-6}$.
 This is of right order as found by COBE. Thus any temperature fluctuations
 above this value would create temperature gradients of such magnitude as to
 move the grains around from high to low density regions and to thereby create a
 uniformity. Only small enough fluctuations would therefore remain intact
 over regions of size $\sim 10^{26}$ cm. Regions of this order are
 characteristic of the curvature size referred to above.

\medskip

The strongest fluctuations will, however, come from rich clusters of
galaxies lying in the present cycle.  These denote the late additions
to the background already in existence from starlight of previous
cycles.
Note that the light from stars belonging to galaxies in clusters of
earlier generations will already have been fully thermalized and such
clusters will not stand up as fluctuations. It is only the newer clusters
 at the last oscillatory minimum with inadequate time to have been
merged in the rest of the thermal background, that would produce fluctuations.
A typical cluster produces the extra starlight, which on
subsequent thermalization produces additional temperature $\Delta T$
over the average background temperature $T$.  Such a fluctuation will
be confined to the neighbourhood of the cluster.  We show next how to
estimate it.
 
Imagine a 10 Mpc size region, typically containing $\approx 10^4$
galaxies, each emitting starlight at the rate $\approx 10^{44}$ erg s$^{-1}$
(corresponding to an average galaxy with absolute magnitude $-$21.4).
The flux of radiation in the form of degraded starlight across the
surface of this region (assumed to be a sphere of radius 5 Mpc), will
be $\approx 3.35 \times 10^{-4}$ erg s$^{-1}$ cm$^{-2}$. Now, the present energy
density of MBR $\approx 4.2 \times 10^{-13}$ erg cm$^{-3}$, implies
that by the $(1+z)^{-4}$ rule, it was $6^4$ times this value at the
epoch of the last minimum of $S$.  Using Stefan's law the flux across
a sphere of radius 5 Mpc for this radiation will be $c/4$ times the
energy density, i.e., about 4.08 erg s$^{-1}$ cm$^{-2}$.  Comparing the
excess flux due to the galaxies in this region computed above with
this value we find that there is an excess flux equal to $8.21\times
10^{-5}$ of the average background.  Equating this to $4\Delta T/T$,
we find that the temperature fluctuation $\Delta T~\cong~56 \mu K$.

Because of the inhomogeneity of the universe on the scale of clusters,
the calculation cannot be made more precise, and at this stage one
should only look at its order of magnitude.  The fact that this
calculation yields a temperature fluctuation of the order reported by
the various observations is, from the point of view of the QSSC, very
encouraging.

Clusters of galaxies are normally associated with a temperature decrement
because of the Sunyaev - Zel'dovich effect.  Indeed observations show such
effects for clusters of redshifts $\lesssim 1$.  By contrast, the effect described
above is predominant for clusters of large redshifts $\sim 5$, corresponding to
the epoch of the last minimum of the scale factor.  Thus we expect the 
thermalization effect to dominate over the S-Z effect for such clusters, 
whereas for later epochs the former effect gets reduced since dust extinction
is much less.  A more detailed study of cluster formation and evolution in the 
QSSC is required to decide at what stage clusters begin to acquire hot gas and when
the gas temperature rises to levels when the S-Z effect becomes important.  
After such a study is carried out it will be possible to estimate the average 
S-Z decrement of temperature in any given direction by integrating over 
redshifts going up to the above stage.   

Over and above the clusters we expect smaller scale fluctuations
on the scales of individual galaxies.  The metallic whiskers are
generated and ejected from inside a typical galaxy.  They have escape
- velocities that take them well beyond the galaxy into its immediate
external environment.  Thus we expect that the extra starlight generated by
the galaxy will provide a slight excess of $\Delta T$ in this region
after thermalization.  The typical length scale of a small group of
galaxies would be $1 - 2$ Mpc.

\section{RELATIONSHIP TO STRUCTURE FORMATION}

From the above discussion it is clear that according to the QSSC the
fluctuations of the MBR arose at the relatively recent epochs
(associated with the last minimum of expansion of the scale factor)
and they would be related to the large scale structure in the
universe.  We will therefore outline briefly the current ideas on this
topic in the framework of the QSSC.  We will consider in particular
the formation and distribution of clusters of galaxies, since they
will turn out to have the most significant effect on the smoothness of
MBR.  For details we refer the reader to Banerjee and Narlikar (1997)
and Nayeri, et al (1999).

\subsection{Gravitational Stability}

To begin with, it is necessary to contrast the structure formation
process in the QSSC with that in the standard cosmology.  In standard
cosmology, structure formation begins in the form of small scale
fluctuations in the spacetime metric as well as matter contents,
fluctuations which are believed to be of quantum origin.  These
fluctuations then evolve as they grow under the effect of gravitation,
with the inflationary phase playing an essential role by modifying the
spectrum of inhomogeneities to the scale-invariant form.  The
inhomogeneities continue to grow with gravitational clustering playing
a major role.  Since radiation and matter (at least the baryonic part)
were strongly coupled in the early epochs, the growth of fluctuations
in matter affects radiation also.  This interaction continues till
the surface of last scattering and the inhomogeneities imprinted on
the radiation background at that stage get imprinted on the MBR and
can be observed today.

It is essential to appreciate that the above scenario {\it does not
  occur} in the QSSC whose past history is quite different from that
of the standard cosmology.  In the QSSC the quantum gravity dominated
epoch as well as inflation did not occur; nor was there a surface of
last scattering.  Thus to understand the presence of fluctuations in
the MBR one needs to look at an entirely different scenario in which
creation of matter at periodic intervals of a finitely oscillating
universe having a long-term de Sitter type expansion, plays a key
role.

That the gravitational force does not play the primary role in
structure formation in this scenario is seen from the work of Banerjee
and Narlikar (1997), wherein these authors looked at the evolution of
small departures from the spacetime metric of the QSSC as well as its
matter density and flow vector.  Specifying the unperturbed metric by
the flat Robertson-Walker line element,

\begin{equation}
ds^{2} = dt^{2} - S^{2}(t)[(dx^{1})^{2} + (dx^{2})^{2} + (dx^{3})^{2}]
\end{equation}

\noindent with $S(t)$ given by (1), the perturbed metric is written as 

\begin{equation}
g_{\mu \nu} = -S^{2} (\eta_{\mu \nu} + h_{\mu \nu}), ~~g_{0\mu} = h_{0\mu}, ~~g_{00} = 1 + h_{00}.
\end{equation}

\noindent Here $\eta_{\mu \nu}$ is the Minkowski metric 
$(\mu, \nu = 1,2,3$; $x^{\mu}$ spacelike coordinates and $x^{0} = t$)
and $\mid h_{ \mu \nu} \mid \ll 1$.  Likewise the density is perturbed
from $\rho_{0}$ to $\rho_{0} + \rho_{1}$, $\mid \rho_{1} / \rho \mid
\ll 1$ and the flow vector from $u_{0}^i= (1,0,0,0)$ to $(u_{0}^{i} +
u_{1}^{i})$ with $\mid u_{1}^i \mid \ll 1$.

The perturbed set of field equations therefore describe the full
gravitational effect on the perturbations.  Banerjee and Narlikar
(op.cit.) found that in a typical oscillation,these perturbations grow
to a limited extent before subsiding.  These authors therefore
concluded that gravitational effects cannot play a major role in
forming large scale structure in the QSSC.

In fact the dynamics of the QSSC requires the existence of an
additional force (besides gravitation) which is derived from a scalar
field $C$ (Hoyle et al, 1995, Sachs et al, 1996).  This scalar field
in turn is related to the creation of matter.
A look at the field equations of the QSSC will help understand why this
field acts in such a way that gravitation does not play a key role in
structure formation. These equations are:

\begin{equation}
R_{ik}-\frac{1}{2} g_{ik} R + \lambda g_{ik}=-\kappa\left[\stackrel{(m)}{T_{ik}} +
\stackrel{(C)}{T_{ik}}\right],
\end{equation}
where $\stackrel{(m)}{T_{ik}}$ is the usual matter tensor while $\stackrel{(C)}{T_{ik}}$
is given by

\begin{equation}
\stackrel{(C)}{T_{ik}}=-f\left[C_i C_k -\frac{1}{4} g_{ik} C^\ell C_\ell \right].
\end{equation}
Both $\kappa=8\pi G/c^4$ and $f$ are positive. Thus the energy tensor of the
$C$-field has a negative coefficient ($-f$) which makes its effect on spacetime
structure repulsive, i.e., acting in direction opposite to gravity. {\it Note
that this tensor acts even when no creation is going on}. This is why, during
a typical oscillation the normal gravitational processes of clustering are
impeded.

A parallel may be cited from standard cosmology with a positive cosmological
constant ($\lambda>0$). In those models which have the $\lambda$-term sufficiently
strong even in the early stages, gravitational clustering and formation of
structures cannot take place. For example, with the presently favoured value of
$\lambda$, clustering ceased at $z \sim 1$.

However, the $C$-field intervenes in a different way through its property that
it induces creation of matter.
As discussed next,
matter is created near collapsed massive objects (i.e., objects close
to becoming black holes) and the created matter is ejected outwards
because of the negative stresses generated by the $C$-field.  It is
this creation phenomenon that holds the key to structure formation in
the QSSC.

\subsection{Matter Creation : A Toy Model}

The field equations of the $C-$field show that at the point of
creation of a particle of momentum $p_i$, the gradient of the
$C-$field satisfies the condition $C_i = p_i$, ($i = 0,1,2,3$).  Thus,
if the rest mass of the particle created is $m_0$, then (with the
speed of light $c = 1$) we have at creation $C_iC^i = m_0^2$.

The use of this relation in the Schwarzschild solution near a massive
object of mass $M$ shows that at a coordinate distance $r$ the
magnitude of the $C$-field energy is raised to

\begin{equation}
C_iC^i =\frac{m^2}{1-\frac{2GM}{c^2r}},
\end{equation}

\noindent where $m^2$ is the magnitude of the energy at large distances from the
massive object.  Now it may happen in general that $m < m_0$, so that
no creation takes place in general. However, near a highly collapsed
massive object for $r$ close enough to the Schwarzschild radius of the
object, the creation condition is satisfied and creation can take
place (Hoyle and Narlikar 1966).

The process normally begins by the creation of the $C$-field along
with matter in the neighbourhood of a compact massive object.  The
former, being propagated by the wave equation, tends to travel
outwards with the speed of light, leaving the created mass behind.
However, as the created mass grows, its gravitational redshift begins
to assert itself, and it traps the $C$-field in the vicinity of the
object.  As the strength of the $C$-field grows, its repulsive effect
begins to manifest itself, thus making the object less and less bound
and unstable. Finally, a stage may come when a part of the object is
ejected from it with tremendous energy. It is thus possible for a
parent compact mass to eject a bound unit outwards.  This unit may act
as a centre of creation in its own right.

We shall refer to such pockets of creation as {\it minibangs} or {\it
  mini-creation events} (MCEs). A spherical (Schwarzschild type)
compact matter distribution will lead to a spherically symmetric
explosion whereas an axi-symmetric (Kerr type) distribution would lead
to jet-like ejection along the symmetric axis.  Because of the
conservation of angular momentum of a collapsing object, it is
expected that the latter situation will in general be more likely.
Using this picture Nayeri et al (1999) proposed a toy model for
structure formation and evolution, which is summarized next.

A large number of points ($N \sim 10^5-10^6$), each one representing a
mini-creation event, is distributed randomly over a unit cubical area
.  The average nearest-neighbour distance for such a distribution will
then be $L\equiv N^{1/3}$.  Now suppose that in a typical
mini-creation event, each point generates another neighbour
point at random within a distance, $d = x L$ in 3$D$ . Here, the
number $x$ is a fraction between $0$ and $1$ .  We shall call $x$ the
separation parameter.  It denotes an ejected piece lying at a distance
$\leq d$ from the parent compact object.
 
The sample cube is then homologously stretched by a linear factor
$2^{1/3}$ to represent expansion of space.  We now have the same
density of points as before, i.e., $2N$ points over a volume of 2
units.  From this enlarged cube remove the periphery so as to retain
only the inner unit cube.This process thus brings us back to the
original state but with a different distribution of an average $N$
points over a unit cube.  This process is repeated $n$ times . Here
the number of iterations, $n$, plays the role of ``time'' as in the
standard models of structure formation.  The number distribution of
points evolves as the `creation + expansion process' generates new
points near the existing ones.

Not surprisingly, soon after, i.e., after $n = 3 - 4$ iterations of
the above procedure, clusters and voids begin to emerge in the sample
volume and create a {\it Persian Carpet} type of patterns. As the
experiment is repeated, voids grow in size while clusters become
denser.  Fig. 1 illustrates the point distribution within a typical
thin slice of
thickness 0.001 of the cube and shows that expansion coupled with
creation of matter is a natural means of generating voids and
clusters.

\begin{figure}
\resizebox{\textwidth}{!}{\includegraphics{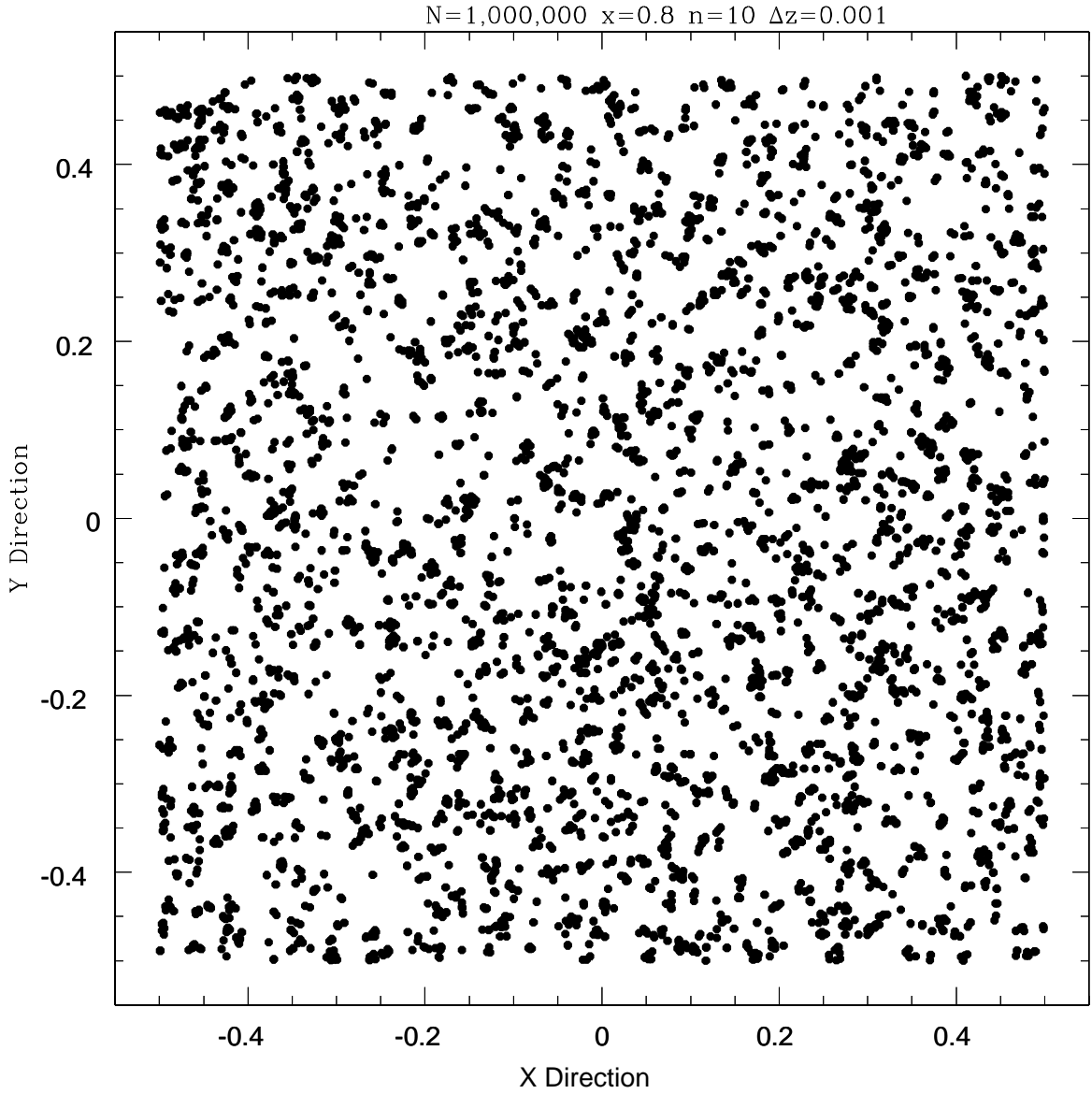}}
\caption{A simulation of large-scale structure in the QSSC, based on the
creation of one generation of clusters in the vicinity of those of the
previous generation, keeping aligned ejection from one generation to the
next. Filaments and voids develop on going from one generation to next.}
\end{figure}

To bring the toy model closer to the reality of the QSSC, Nayeri, et
al proceeded as follows. Since the creation activity is expected to be
confined largely to a narrow era around a typical oscillatory minimum,
when the $C$-field is at its strongest, by considering the number
density of collapsed massive objects at one oscillatory minimum of
QSSC to be $f$, the number density at the next oscillatory minimum
would fall to $f \exp{(-3Q/P)}$, that is, {\it if no new massive
  objects were added}. Thus to restore a steady state from one cycle
to the next, within each unit volume

\begin{equation}
\omega f \equiv [1 - \exp{(-3Q/P)}] f \sim (3Q/P) f,
\end{equation}

\noindent masses must be created anew. In other words, a fraction $\omega$ of
the total number of massive objects must replicate themselves in the
above fashion.

Notice that, unlike the old steady state theory which had new matter
appearing continuously, we have here discrete creation, confined to
epochs of minimum of scale factor. The `steady-state' is maintained
from one cycle to next.  Which is why the above fractional addition
$\omega$ is required at the beginning of each cycle.

Therefore, instead of creating a new neighbour point around each
and every one of the original set of $N$ points, one does so only
around $\omega N$ of these points chosen randomly, where the fraction
$\omega$ is as defined in equation (11).  Likewise, the sample volume
is homologously expanded by the factor $\exp{(3Q/P)}$ only instead of
by factor 2.

\begin{figure}
\resizebox{\textwidth}{!}{\includegraphics{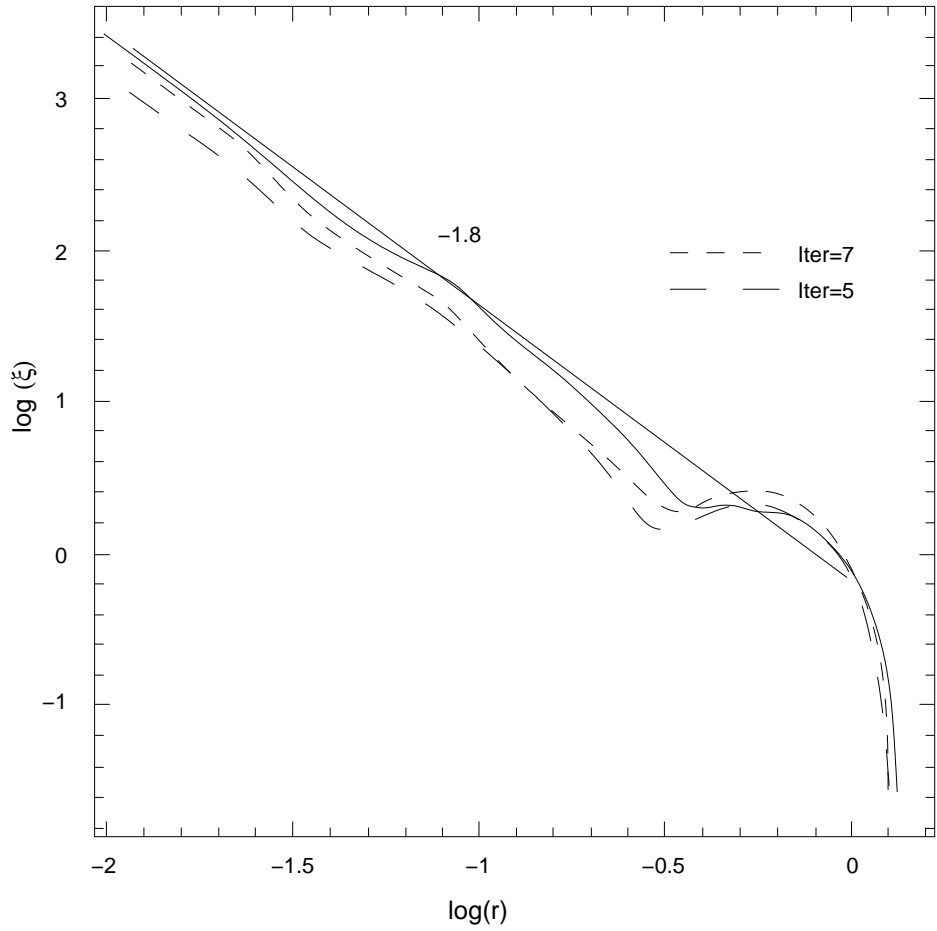}}
\caption{A two-point correlation function for simulations of the kind
shown in Fig. 1 has a power-law-type distribution with index approaching
$- 1.8$ after a few iterations of the process of creation. The evolution of
the index towards the ideal $- 1.8$ slope line (shown by the continuous line)
is shown as the process goes on for up to ten generations.}
\end{figure}

 \section{The Two Point Correlation Functions}

Although visual inspection of Figures like 1 suggests
that the toy model is proceeding along the right lines, a {\it
quantitative} measure of the cluster-void distribution helps in
comparing simulations with reality.
The dimensionless autocorrelation function

\begin{equation}
\xi (r) = <[\rho ({\bf r})-<\rho>][\rho ({\bf r}_1 + {\bf r})-<\rho>]>/<\rho>^2,
\end{equation}

\noindent where $<\rho>$ is the average density in the volume,
is one convenient measure of such irregularities in the space
distribution.  Typically, different classes of objects cluster at
different characteristic lengths.  To fix ideas Nayeri et al (1999)
looked at distribution of clusters of galaxies.  Observationally, it
is believed that the two point correlation function for cluster
distribution obeys the following scaling law:

\begin{equation}
\xi _{cc}(r) = \left(\frac{r}{r_0}\right)^{-\gamma},
\end{equation}

\noindent with $\gamma \simeq 1.8$ and $r_0 = 25h^{-1}$ Mpc, where the
Hubble constant at the present epoch is taken to be $100h$
kms$^{-1}$Mpc$^{-1}$. In order to quantify the issues of formation of
structures in this scenario Nayeri, et al (op.cit) took the following
measures.

It is known that instead of having a uniform distribution of matter on
large scales, the observed universe has structures of typical sizes of
a few tens of megaparsecs .  These ``structures'' are regions of
density considerably higher than the background density, with the
maximum density contrast $\delta = (\delta \rho (r) /< \rho >)$ going
from order unity (in the case of clusters) to a few thousand (in the
case of the galaxies).

Any process which generates structures must be able to produce to the
zeroth order, entities whose density contrast is of such magnitude and
with the property that on larger and larger distance scales, the
density contrast became less significant. This ensures that on a large
enough scale the universe is homogeneous.

Given this prescription for generating structures without
gravitational dynamics, Nayeri, et al first ensured that the visual
impression created by the cluster simulations did imply that as the
number of iterations were increased the number of high density regions
also increased. In the initial (totally random) configuration one expects to
find regions of high density arising only because of the Poisson noise. In
the later ``epochs'' after a few iterations, however, one would expect
to find, as in a clustering scenario, that the variation of the one
point distribution function for density $(\rho/< \rho >)$ with
$<\rho>$ the average density in the volume, showed a steady and
significant increase in the the number of high and intermediate
density regions. This expectation was borne out and it was also
observed that the value of maximum density also increased as a
function of the number of iterations, which in this experiment
corresponds to ``time''. The density field was generated on a grid
placed into the simulation volume using the algorithm of cloud in
cell.  The simulations showed the growth of structures through rise in
the density maximum as a function of number of iterations.

A quantitative measure that was computed from these data sets was the
two-point correlation function.  Fig. 2 shows the
two point correlation function for the case of the QSSC based model.
As ``time'' goes on, the slope of the correlation function gets closer
and closer to $-1.8$.  From the value of the X-axis intercept of the
two point correlation function Nayeri, et al could get a rough
estimate of the size of the structures in units of the size of our
simulation box. From their results they estimated that the size of the
structures formed is approximately $\beta =0.15-0.3$ times the
boxsize. If one sets these values equal to the observationally
accepted value of $r_0$, one can get a better physical sense of the
results. If we set, $\beta= 0.3$, say, and $r_0 = 25 h^{-1}$ Mpc, then
the linear size of the simulation box would be $ \sim 84 h^{-1}$ Mpc.
Typical cluster masses and sizes resulted from such a numerical
exercise.

This may be seen as an attempt to relate the toy model to a realistic
cosmological scenario. Of course, as the above exercise shows, the
results can be scaled up/down by rescaling the simulation parameters
and thus are independent of the `absolute' size of the box. A more
detailed dynamical theory of the creation process will help
relate the absolute size of clustering to the theoretical parameters.

We now examine the typical angular scales of these three types of
inhomogeneities and how they are reflected in the angular power
spectrum of the MBR. So far as clusters are concerned, we will fold in
the above evolving two-point correlation function into our calculation.

\section{THE ANGULAR POWER SPECTRUM OF INHOMOGENEITIES}

Typically we will look at the angle subtended by a linear scale $a$ of
an object located at redshift $z$ in the quasi-steady state model
whose scale factor is defined in equation (1). To compute the above angle,
we define the dimensionless parameters by the
following formulae:

\begin{eqnarray} 
\Omega_0&=&\frac{8\pi G\rho_0}{3H_0^2}~~~{\rm density~  parameter},\nonumber \\
\Lambda_0&=&\frac{\lambda}{3H_0^2}~~~
{\rm cosmological~ constant~ parameter}, \nonumber
\\ \Omega_{c0}&=&\frac{8\pi G\rho_{c0}}{3H_0^2}~~~{\rm creation~ density~ 
parameter}, \nonumber \\ 
K_0&=&\frac{k}{H_0^2S_0^2}~~~{\rm curvature~ parameter}.
\end{eqnarray}

\noindent Here $\lambda$ is the cosmological constant, which is negative in the
QSSC.  In the model considered by Banerjee et al (2000) as well as
here, $K_{0} = 0.$ The angle subtended at the observer by the above
scale at redshift $z$ ($see$ Banerjee and Narlikar, 1999) is then
given by

\begin{equation}
\alpha = \frac{H_0a(1~+~z)}{c}\left[\int^{1+z}_{1}\frac{dy}{\sqrt{\Lambda_0~+~\Omega_0y^3~+~\Omega_{c0}y^4}}\right]^{-1}.
\end{equation}

\noindent We will take $z~=~z_{{\rm max}}~=~5$.  (The maximum redshift used here is 
indicative only.  The results do not change much even if it were
increased to 6-10.)  The calculation, using the parameters of (5), then gives

\begin{equation}
\alpha~= \frac{\pi a}{3600}.
\end{equation}

\noindent 
where $a$ is measured in units of megaparsecs. Unless explicitly
stated otherwise, angles will be referred to in radians.

\medskip
\noindent 
This then is the characteristic angle subtended by an object of linear
size $a$ Mpc, viewed when close to the last minimum.  Assuming
that this is the size of a rich cluster of galaxies, we expect that
the characteristic angular size of the MBR inhomogeneity it generates
will be of this order.  As we will shortly show, this does translate
to a peak in the angular power spectrum at a harmonic of order $l
~\approx 1/\alpha$. So we arrive at the following approximate
relationship between cluster size and the harmonic it generates:

\begin{equation}
l~\cong \frac{3600}{\pi a}.
\end{equation}

If we set $a {\rm ~Mpc~} \approx 2 \times$ radius of curvature at the
last minimum, we expect a large angle peak at $l \approx 6-10$.  The
smaller angle peaks will occur at much larger values of $l$.  Thus a
peak in the power spectrum of MBR anisotropy at harmonic $l~\approx
200$ as observed by the Boomerang and the Maxima groups (de Bernardis
et al 2000, Hanany et al, 2000) is generated in this theory by rich
clusters of diameter $\approx 5.5$ Mpc, located at cosmological
distances.  Smaller peaks are expected on the scale of groups of
galaxies, at the higher values of $l~\approx 500-1000$, and on the scale
of superclusters at the lower values $l~\approx 10-20$.  These latter peaks will be
smaller than the cluster peak at $l ~\approx 200$ and thus they will
be much harder to detect.

We next show that these expectations are borne out by detailed
calculations of this effect on the angular power spectrum of MBR.  We
model the above effect along the lines of Hajian and Souradeep (2002) 
as follows.  Imagine a typical type of inhomogeneity as a
set of small disc-shaped spots, randomly distributed on a unit
sphere.  The spots may be either `top hat' type or `Gaussian'
type.  In the former case they have sharp boundaries whereas in the
latter case they taper outwards.  We will assume the former for
clusters, and the latter for the curvature effect and also for
galaxies or groups of galaxies.  This is because the clusters will
tend to have rather sharp boundaries whereas in the other two cases
such sharp limits do not exist.

Let us begin with the simple case of a random distribution of
infinitesimally small spots. With each direction $\hat{n}_i$ for a
typical spot centre on the sky chosen randomly, we have from $N$ such
spots of roughly identical temperature fluctuation $\Delta T_c$

\begin{equation}
\Delta T(\hat{n}) = \sum^{N}_{i=1} \hat{a}_{n_i} 
\delta (\hat{n} - \hat{n}_{i})\,\, \Delta T_c\,\,.
\end{equation}

\noindent Here $\hat{n}$ is a general direction on the unit sphere.  
If the locations of the spots in the sky are uncorrelated, we have

\begin{equation}
\langle\hat{a}_{n_i} ~ \hat{a}_{n_j}\rangle = \delta_{ij}.
\end{equation}

\noindent   Writing $\Delta T$ as a function on the sphere, we expand it in a
series of $Y_{lm} (\theta, ~\phi)$:

\begin{equation}
\Delta T(\hat{n}) = \sum_{lm} a_{lm}Y_{lm}(\hat{n}).
\end{equation}

\noindent The expansion of $\delta (\hat{n} - \hat{n}_{i})$ in terms of the 
spherical harmonic functions allows us to obtain

\begin{equation}
a_{lm} =  \sum^{N}_{i=1} \hat{a}_{n_i}Y^{\star}_{lm}(\hat{n}_{i}) \,\,\Delta T_c\,.
\end{equation}

\noindent The corresponding angular power spectrum is given by

\begin{equation}
{\mathcal C}_{l} \equiv \frac{l(l+1)}{2\pi}\,\,\frac{1}{(2l+1)} \sum_{m} 
\langle a^{\star}_{lm} ~a_{lm}\rangle  = \frac{l(l+1)}{2\pi} \frac{N}{4 \pi} (\Delta T_c)^2.
\end{equation}

\noindent 
In the last step we use eq.~(19) that encodes our simplistic
assumption of uncorrelated spots.  We get, for large $l$, ${\mathcal
  C}_{l}~ \propto ~ l^{2}$.  This result is valid for angular scales
much larger than the angular scale of the typical spot. The $l^2$ rise
is a direct consequence of assuming that the
location of the spots are uncorrelated. More detailed modeling where
the correlation between the spots is included would serve to soften
the $l^2$ rise. It may be an important issue when comparing our
results with observations of MBR inhomogeneity on large angular
scales.

However, we need to model spots of finite sizes ranging from $\approx
10^\circ$ to a few arc minutes, corresponding to the three typical
inhomogeneities discussed above.  For spots of finite size, assumed to
be circularly symmetric with a given profile centered on the typical
spot centre, we have a modification of (18) in the form
(where the delta function is replaced by a function $f$):

\begin{equation}
\Delta T(\hat{n}) = \sum^{N}_{i=1} \hat{a}_{n} \,
\Delta T_c \, f(\hat{n}~\cdot~\hat{n}_{i}).
\end{equation}

\noindent Using the circular symmetry of the spots, we expand

\begin{equation}
f(\hat{n}\cdot\hat{n}_{i})= \sum_{l} \frac{(2l+1)}{4 \pi}
f_{l}~P_{l}(\hat{n}\cdot\hat{n}_{i}),
\end{equation}

\noindent in terms of its Legendre transform, $f_l$.  
It is then possible to express 

\begin{equation}
{\mathcal C}_{l} =  \frac{l(l+1)}{2\pi} \frac{(\Delta T_c)^2}
{4 \pi} \sum_{i}  \sum_{j}<\hat{a}_{n_i} \hat{a}^{\star}_{n_j}>
f^{2}_{l}P({\hat n}_{i}\cdot  {\hat n}_{j}).
\end{equation}

\noindent Assuming the location of the $N$  spots to be  uncorrelated 
and randomly distributed (see eq.~(19)) this gives

\begin{equation}
{\mathcal C}_{l} =  \frac{l(l+1)}{2\pi}  \frac{N}{4 \pi}(\Delta T_c)^2  f_l^2.
\end{equation}

 For a normalized Gaussian spot profile with typical (1-$\sigma$ width)
angular size $\alpha$, the function $f_l$ is given by

\begin{equation}
f_{l} (\alpha) = {\rm exp}~(-l^{2}\alpha^{2}/2).
\end{equation}
\noindent The angular power spectrum ${\mathcal C}_{l} \propto 
l^{2}e^{-l^{2}\alpha^{2}}$ has a single peak at $l_p \approx
\alpha^{-1}$.


For a top hat profile, wherein the spot has uniform temperature
fluctuation across a finite disc of angular size $\alpha$, with a
sharp drop-off to the background temperature outside, the function
$f_l$ is given by

\begin{equation}
f_{l} (\alpha) =  \frac{1}{(l+1)^2}
\left[\frac{\cos \alpha ~ P_l(\cos \alpha)-P_{l-1}(\cos  \alpha)} 
{ (2 \sin^2\alpha/2)}\right]\,\,.
\end{equation} 
The ${\mathcal C}_{l}$ in this case has a series of peaks in the
multipole space. The first peak in ${\mathcal C}_l$ for top-hat spots
occur at $l_p \approx \pi/2\alpha$ and has the largest amplitude. The
successive peaks occur roughly at integer multiples with diminishing
peak amplitude.

We therefore consider a composite model in which the inhomogeneities
of the MBR arise from a superposition of random spots of three
characteristic sizes corresponding to the three effects discussed
above.  From the order of magnitude estimates made earlier, we expect
three peaks in the angular power spectrum at their respective $l$ -
ranges.  To check the accuracy of this expectation, we consider a $6$
- parameter modeling of the angular power spectrum

\begin{eqnarray}
{\mathcal C}_{l} &=& A_1 \,\,~ l(l+1) e^{-l^{2}\alpha^{2}_{1}} + 
A_2\,\, \frac{l}{l+1}\left[\frac{{\rm ~cos~} \alpha_{2} P_{l} ({\rm cos}~ \alpha_{2}) 
- P_{l-1}({\rm cos}~ \alpha_{2})}{2 \sin^2\alpha_2/2}\right]^2 \nonumber \\ {}&& + A_3\,\, l(l+1)e^{-l^{2}\alpha^{2}_{3}}
\end{eqnarray}
\noindent with the parameters $\{A_1 , A_2, 
A_3,\alpha_{1},\alpha_{2},\alpha_{3}\}$ determined by obtaining the
best fit with the present observations of the angular power spectrum
of MBR inhomogeneities. The parameters $A_1$, $A_2$ and $A_3$ depend
on the number density as well as the typical temperature fluctuation
of each kind of spot. However, given $A_1$, $A_2$ and $A_3$, it is
possible to compute the {\it rms} temperature fluctuation contributed
by each component to the overall MBR temperature fluctuations.

If the model is successful, the values of $\alpha_{1}, \alpha_{2},
\alpha_{3}$ (or, the corresponding multipole value $l_p$ at which the
${\cal C}_l$ from each component peaks,) and the {\it rms} temperature
fluctuation for components would turn out to be within the range of
values expected on the basis of the physics of the processes.

The analysis till now has assumed that the distribution of hot spots is
uncorrelated. As described in the previous section, toy model of
structure formation within QSSC does recover the observed power law
spatial correlation in the distribution of clusters. We extend the
computation of CMB anisotropy spectrum to the case when the hot spots
are correlated. We do this only for the hot spots that are linked to
the clusters.  Then eq.~(19) is revised to
\begin{equation}
\langle \hat{a}_{{\bf n}_i} \hat{a}_{{\bf n}_j}^{*}
\rangle\, =\, w(\theta)
\label{spotcorr}
\end{equation}
where $w(\theta)$ is angular correlation of clusters on angular scale,
$\theta= \cos^{-1}({\bf n}_i.{\bf n}_j)$.  The form of ${\mathcal
  C}_l$ in eq.~(26) is revised to
\begin{equation}
{\mathcal C}_l = \frac{A N}{8\pi^2} l(l+1) f_l^2 u_l.  
\end{equation}
where $w(\theta) = \sum_l (2l+1)/(4\pi) u_l P_l({\bf
  n}_i.{\bf n}_j)$.  The angular correlation $w(\theta)$
can be related to the $3$-D spatial correlation function $\xi(r)$
using the well known Limber equation ( $see$ Peebles, 1980). In
particular, for a power law correlation function $\xi(r) =
(r/r_0)^{-\gamma}$, it can be shown that $w(\theta) \propto
\theta^{1-\gamma}$ for $\theta\ll 1$ limit and $1<\gamma<6$.  In this
case, $u_l \propto l^{\gamma-3}$ for $\gamma \le 3$ (Peebles and
Hauser, 1974).  The correlation of the hot spots due to clusters are
expected to be bounded within $\gamma=3$, the limit of uncorrelated
spots and $\gamma \sim 2$, the correlation function that the clusters
distribution evolves to within the QSSC model.  Consequently, we also
consider an extended model with an extra parameter $\gamma$,
corresponding to an angular power spectrum
\begin{eqnarray}
{\mathcal C}_{l} &=& A_1 \,\,~ l(l+1) e^{-l^{2}\alpha^{2}_{1}} +
A_2\,\, \frac{l^{\gamma-2}}{l+1}\left[\frac{{\rm ~cos~} \alpha_{2} P_{l} ({\rm cos}~ \alpha_{2})
- P_{l-1}({\rm cos}~ \alpha_{2})}{2 \sin^2\alpha_2/2}\right]^2 \nonumber \\ {}&& + A_3\,\, l(l+1)e^{-l^{2}\alpha^{2}_{3}}
\end{eqnarray}
where all the other parameters are as described in eq.~(29).

\begin{figure}
\resizebox{\textwidth}{!}{\includegraphics{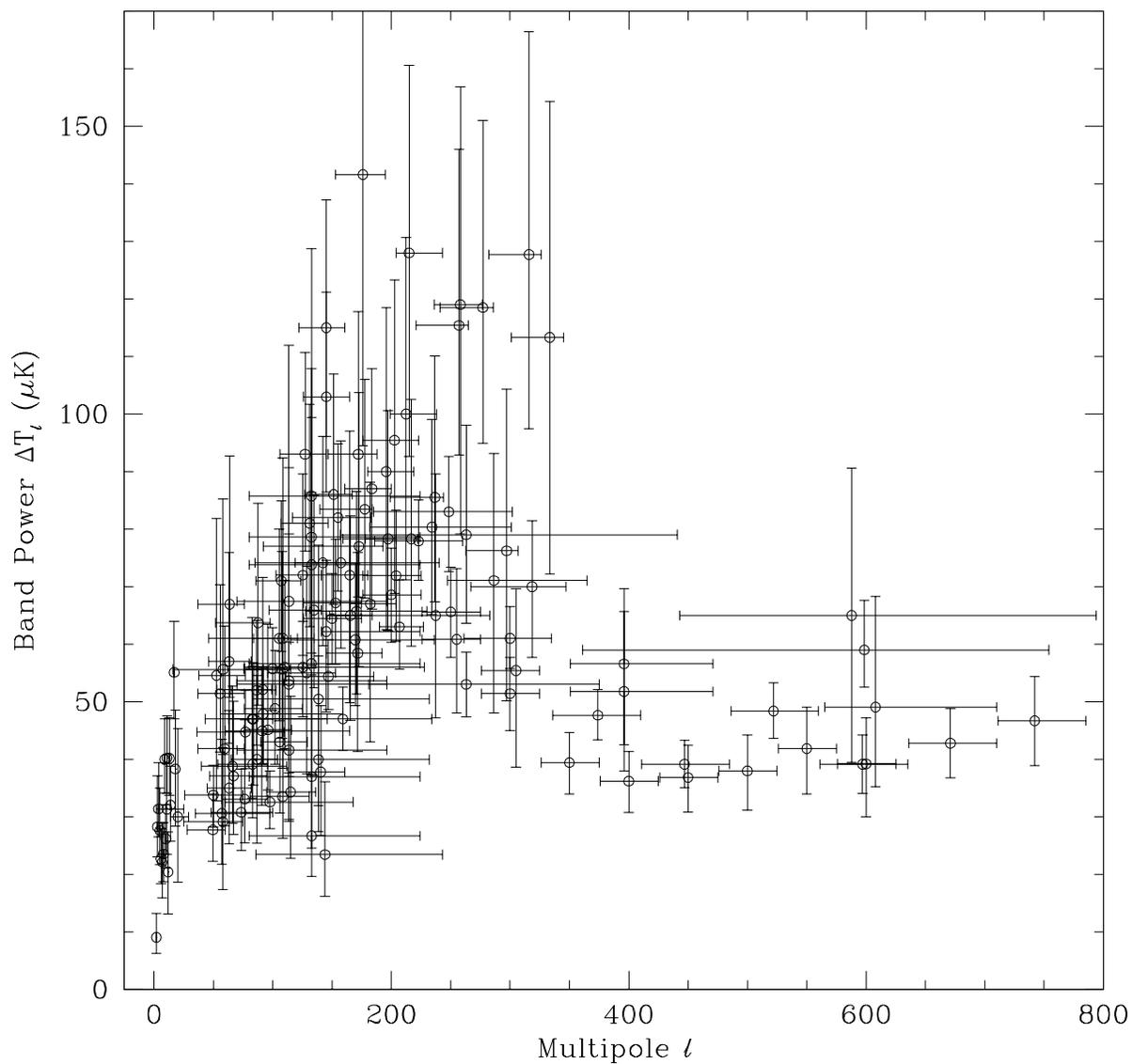}}
\caption{The band power spectrum of the MBR
temperature inhomogeneity, measured ($> 2\sigma$ detections only) by
the different experiments. The compilation of Podariu et al. 2001 has
been used.}
\end{figure}

\section{COMPARISON WITH OBSERVATIONS}

We now compare our `three component model' with a published data
compilation available at the time of writing.  Fig. 3 describes the
error-bars of the observed data which indicates the extent of scatter
at present (Podariu et al.  2001).  The data have been binned into
sixteen bins in multipole space with a mean $\Delta
T_l\equiv\sqrt{{\mathcal C}_l}$ over the bin and a $1-\sigma$
error-bar (Podariu et al. 2001, table 2).  Two peaks are easily
visible at $l \sim 10$ and 200 while a third peak at the much higher
value of $\sim 600$ is probably present.  The $l$ - values can be
related to physical dimensions of the sources of inhomogeneities by
the formula (17). For the actual fitting, we have used the above binned
data.

We first discuss the case of uncorrelated spots ($\gamma = 3$).
Fig. 4 shows the `best-fitting' angular power spectrum curve $\Delta
T_l$ obtained for QSSC by using a six parameter model for the
characteristic angular sizes, $\alpha_i$ of the spots and the
corresponding amplitudes, $A_i$. For a given type of spot, the angular
size can be readily related to the multipole $l_p$ where the
contribution to ${\mathcal C}_l$ peaks.  The amplitudes are used to
compute the {\it rms} temperature fluctuation, $\Delta T_{\rm rms}$,
contributed by each of the components. For the `best fit' model we
find $l_{p1} \cong 27$, $l_{p2}\cong 182$, $l_{p3}\cong 502$, $\Delta
T_{\rm rms1}\cong 20 \mu K$, $\Delta T_{\rm rms2}\cong 31 \mu K$ and
$\Delta T_{\rm rms3}\cong 16 \mu K$.  For the QSSC model we have used
six parameters to obtain the above fit with a minimum $\chi^2$ of 16.8.
We divide the minimum
$\chi^{2}$ value by $16-6=10$ degrees of freedom to arrive at the
value $1.68$ for the reduced $\chi^2$ for our fit.
This is useful for a comparison with any other model.

It is encouraging to note that the parameters for the best fit model
are broadly consistent with our expectations. The value of $l_{p3}
\cong 500$ is somewhat below the expected range of $600$ -- $900$.
This is most likely because the largest central multipole value of the
binned data set is $l\sim 600$.  As MBR anisotropy on large multipole
bands are determined and included in the binned data, we expect the
best fit values of $l_{p3}$ (and possibly, $\Delta T_{\rm rms3}$) to
go up.  We also note the significance of the top-hat approximation to spot
profile in clusters in our fit to the data. If we replace the top-hat
profile of cluster anisotropy by a Gaussian then we can find a best
fit model with $\chi^2\sim 23$ with parameters $l_{p1} \cong 25$,
$l_{p2}\cong 201$, $l_{p3}\cong 900$, $\Delta T_{\rm rms1}\cong 19 \mu
K$, $\Delta T_{\rm rms2}\cong 30 \mu K$ and $\Delta T_{\rm rms3}\cong
20 \mu K$. Thus the reduced $\chi^2\sim 2.3$.
The similarity in the amplitude and sizes clearly
demonstrates the robustness of our model to finer details of the spot
profile. The reality may lie between the two limits, the Gaussian profile
and the top-hat one.

\begin{figure}
\resizebox{\textwidth}{!}{\includegraphics{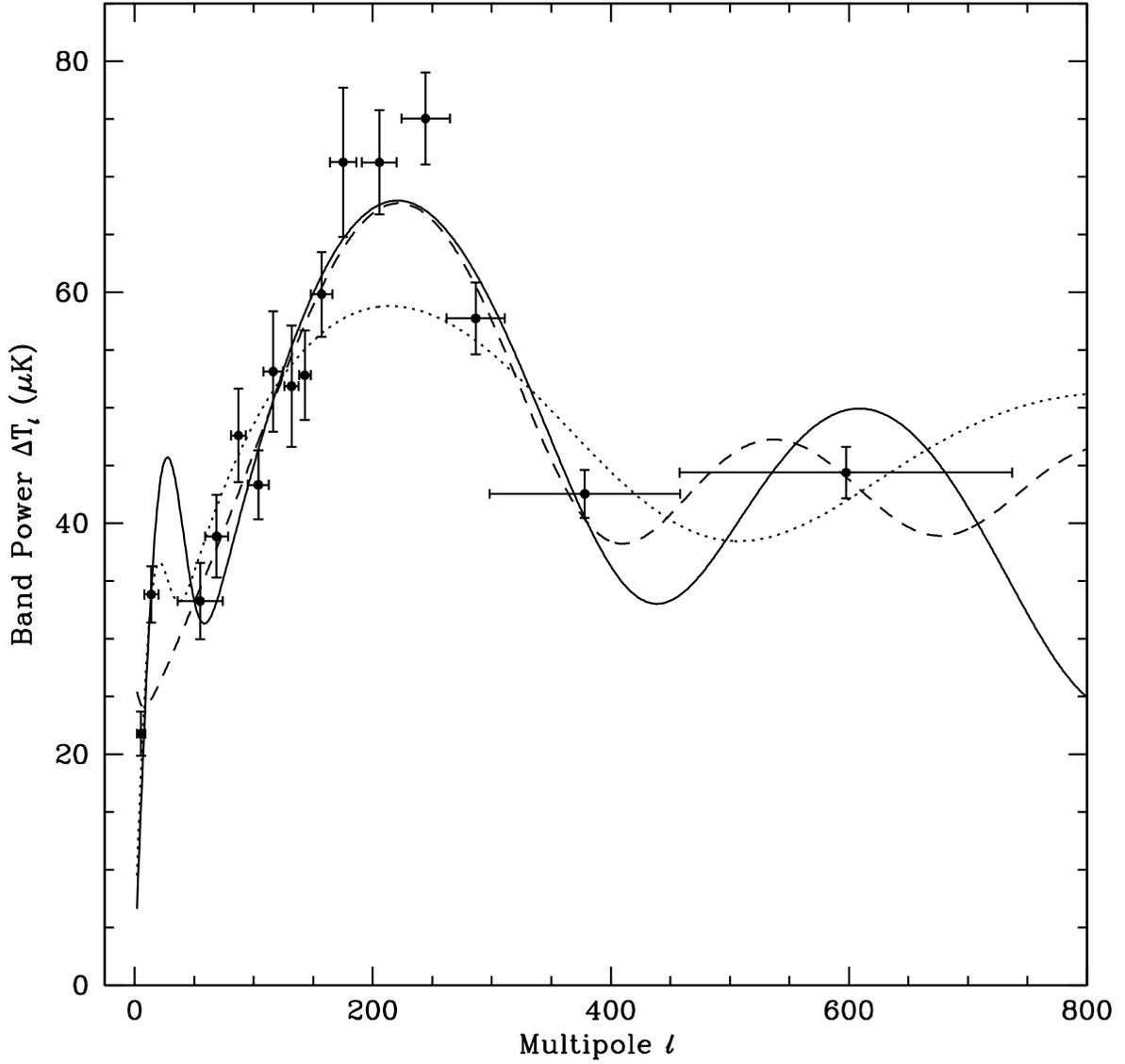}}
\caption{Best-fitting angular power spectrum
curves in the QSSC for $\gamma=3$ (continuous curve) and $\gamma=2$
(dotted curve) are plotted and are compared with the one for the
favoured big bang model with $\Omega_\Lambda=0.7$ (dashed curve).
The fitted data set is the one described in Figure~3 and has been
averaged into 16 bins in multipole space (Podariu et al. 2001).}
\end{figure}

\begin{figure}
\resizebox{\textwidth}{!}{\includegraphics{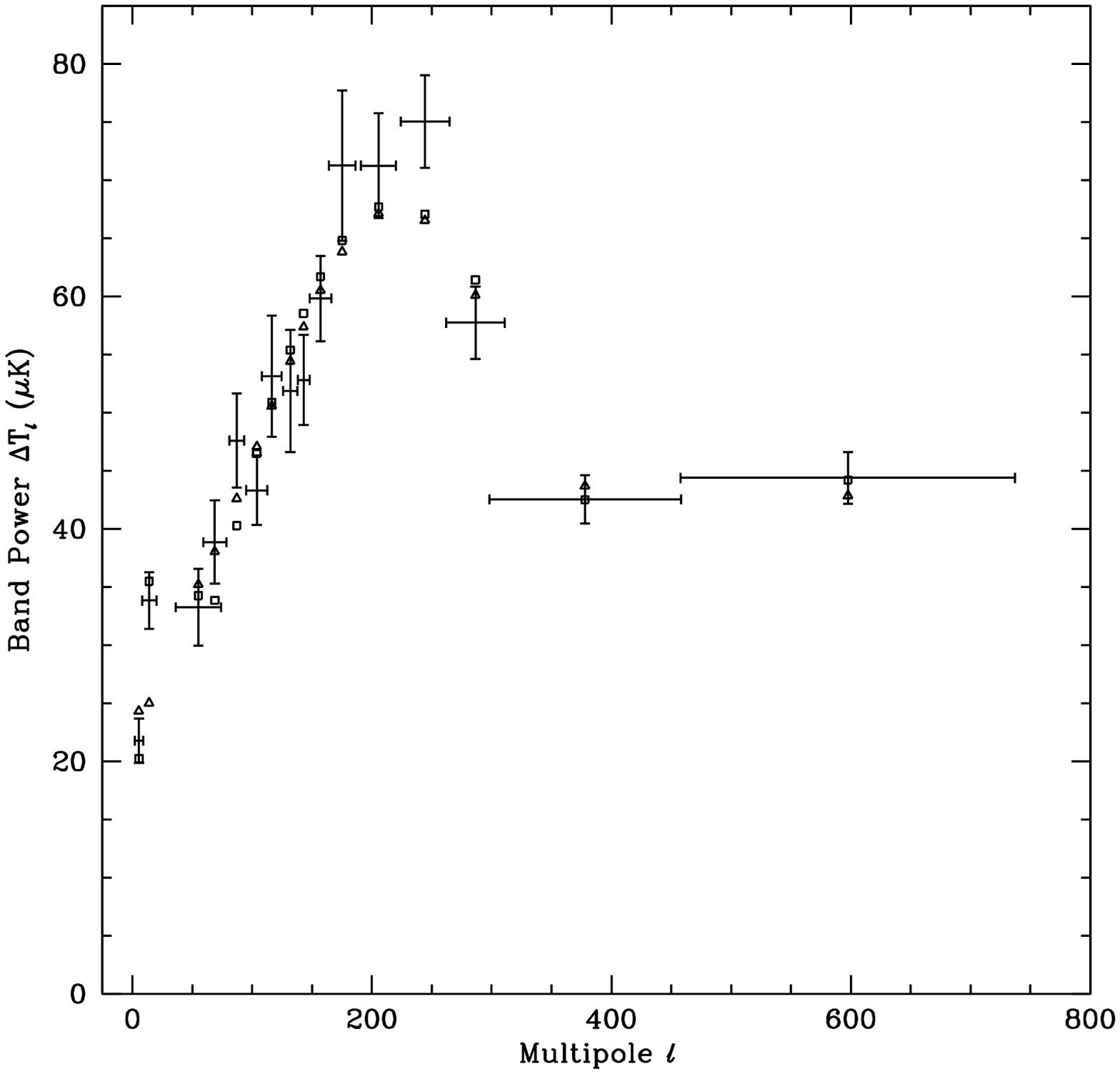}}
\caption {The predicted band power $\Delta T_l$ values
(square root of ${\mathcal C}_{l}$ averaged over the $16$ multipole bins), are
computed for the best-fitting $\gamma=3$ QSSC model (shown by squares)
and the favoured big bang model (shown by triangles)
and are compared with the observed binned values.}
\end{figure}

We compared with the same binned data, the anisotropy spectrum
prediction of a grid of open-CDM and $\Lambda$-CDM models within the
standard big bang cosmology. We varied the matter density, $\Omega_0
=0.1$ to $1$ in steps of $0.1$; the baryon density, $\Omega_b h^2$
from $0.005$ to $0.03$ in steps of $0.004$ where $h$ is the Hubble
constant in units of $100$ km s$^{-1}$ Mpc$^{-1}$; and the age of the
model, $t_0$ from $10$ Gyr to $20$ Gyr in steps of $2$ Gyr (Sugiyama,
1995).  For each value of $\Omega_0$ we considered an open model and
one where a compensating $\Omega_\Lambda$ was added to get a flat
model. For the same binned data set, we found the minimum value of
$\chi^2=23$ for the flat universe model with $\Omega_0=0.7$
($\Omega_\Lambda=0.3$), $\Omega_b h^2=0.021$, $t_0=14$ Gyr ($h=0.52$).
Recent MBR inhomogeneity data, when combined with high redshift
supernovae data that support non-zero $\Omega_\Lambda$ and constraints
on $\Omega_0$ from large scale structures observations, favour flat
cosmological models with higher values of $\Omega_\Lambda$. In our
analysis, these models have higher, but comparable, $\chi^2$ values,
eg., for $\Omega_b h^2=0.021$ and $t_0=14$ Gyr, the models with
$\Omega_\Lambda=0.6$ ($h=0.62$) and $\Omega_\Lambda=0.7$ ($h=0.67$)
have $\chi^2 = 24$ and $25$, respectively.  For a comparison, the
corresponding curve for the favoured big bang model
($\Omega_\Lambda=0.7$) is also shown in Fig. 4.

Next we incorporate the correlation in the spots arising from clusters
corresponding to spatial correlation in the distribution of clusters.
In the absence of a detailed modeling of CMB anisotropy arising from
clusters, we admit the entire range $3\ge\gamma\ge 1.8$.  For
$\gamma\sim2$, we find the best fit values $l_{p1} \cong 18$,
$l_{p2}\cong 230$, $l_{p3}\cong 977$, $\Delta T_{\rm rms1}\cong 13\mu
K$, $\Delta T_{\rm rms2}\cong 32 \mu K$ and $\Delta T_{\rm rms3} \cong
23\mu K$.  For these best fit parameters, $\chi^{2} \cong 39$. As seen in
Fig. 4, most of the discrepancy arises at the observed peak around
$l\sim200$, particularly from the 13th point whose contribution to
$\chi^{2}$ alone is $\approx 37$ percent. This point also does not fit the
standard big bang models and seems to be a general {\it outlier}.
With $\gamma=2$ the second peak in our model is broad and
cannot fit the high amplitude of the peak observed in the binned
compilation.  At present, this may not be a significant problem for our
model even with $\gamma=2$. As $\gamma$ is increased to $3$, the fit
improves rapidly. We are not in a position to fix the value of
$\gamma$ that would be appropriate for the clusters at $z\sim5$ within
QSSC. On the observational front, the amplitude at the peak needs to
be well established by an experiment such as MAP that covers both the
low $l$ and high $l$ parts with good $l$-space resolution.  The plot
of the individual band power estimates in Fig. 3, shows
considerable dispersion in the central value of band power estimates
and some of discrepancies are attributed to calibration uncertainty.

In Fig. 5, we compare the band power values $\sqrt{{\mathcal C}_{l}}$, averaged
within the $16$ multipole bins for the best fit QSSC and the favoured
big bang ($\Omega_\Lambda=0.7$) models, with the binned values of the
observed MBR inhomogeneity data. Within QSSC, the unclustered
hot-spots allow for a very good fit to the CMB data up to $l ~ 600 $.
The peak around $l\sim 200$ in the present data is best fit by
uncorrelated spots and fit around the peak is adversely affected as
the correlation between the spots is increased.  We also find
that the details of spot profile does not affect the fit
significantly.


\section{CONCLUDING REMARKS}

In the QSSC the MBR arises from the matter in galaxies and this is
quite different from its origin in the hot big bang cosmology. For
example, the major peak (at $l \approx 200$) in the MBR in the
framework of the QSSC is explained in terms of rich clusters of
galaxies, whereas in the hot big bang cosmology it is the `Doppler'
peak associated with acoustic oscillations of the photon -baryon
fluid. The QSSC interpretation links the inhomogeneities of the
radiation field to those of the matter field, visible matter in the
form of clusters of galaxies.  Indeed, one may argue that in the QSSC
interpretation, the observations of MBR-inhomogeneities provide us
with a direct diagnostic of the structural hierarchy in the universe.
For example, the pattern of patchiness around $l\sim 200$ depends on the
cluster - cluster correlation function.  However, we must use the
clusters at large redshifts ($\approx 5$) for accurately estimating
the magnitude
of the effect. In this paper we account for effect of cluster-cluster
correlation function based on earlier published results obtained using
a `toy model' structure formation model within QSSC.  We hope to carry
out such an analysis when more complete data on the cluster
distributions is available going back to epochs of high redshifts and
the theory of structure formation and evolution in the QSSC makes
further progress.

Additionally, from Fig. 4 we see that peaks occur at different
locations for $l>500$, for the most favoured big bang model, and the
best-fitting QSSC model. Future observations at high values of $l$
from MAP, PLANCK and other experiments may help distinguish between
the two cosmologies.

In the end perhaps it is best to stress the attitudinal difference
between the two cosmologies. In the big bang cosmology, the inferences
are related to the postulated initial conditions prevailing well beyond
the range of direct observations (at redshifts $\gtrsim$ 1000). In the
QSSC the attempt is to relate patchiness of structure, (at redshifts
$\approx 5$) which may be observable one day, to the patchiness of MBR.
These latter studies admittedly do not give predictions as sharp as
those given by the former, but they may perhaps claim to be less
speculative.

\vspace{1cm}\noindent
{\bf Acknowledgements}

J.V. Narlikar and R.G. Vishwakarma acknowledge grants from the
Department of Atomic Energy for support of the Homi Bhabha Distinguished
Professorship for the former and the associated postdoctoral
fellowship for the latter.  G. Burbidge wishes to acknowledge the
hospitality given to him during his visits to IUCAA in 2001-2002.  A.
Hajian thanks IUCAA for supporting him as a project student for six
months.


\centerline {{\bf References}}
\medskip

\noindent Banerjee, S.K. and Narlikar, J.V. 1997, {\it Ap. J.}, {\bf 487}, 69
\medskip

\noindent Banerjee, S.K. and Narlikar, J.V. 1999, {\it MNRAS}, {\bf 307}, 73
\medskip

\noindent Banerjee, S.K., Narlikar, J.V., Wickramasinghe, N.C., Hoyle, F. and Burbidge, G. 2000, {\it A.J.}, {\bf 119}, 2583
\medskip

\noindent Burbidge, G. and Hoyle, F. 1998, {\it Ap. J.}, {\bf 509}, L1
\medskip

\noindent de Bernardis P., et al 2000, {\it Nature}, {\bf 404}, 955
\medskip
 
\noindent Dittmar,W. and  and Neumann, K. 1958, in {\it Growth and Perfection in Crystals}, Eds R.H. Doremus, P.W. Roberts and D. Turnbull, Wiley,388
\medskip

\noindent Gomez, R. J. 1957, {\it Chem. Phys.}, {\bf 26}, 1333
\medskip

\noindent Hajian, A. and Souradeep, T., {\it in preparation}, 2002
\medskip

\noindent Hanany, S. et al 2000, {\it Ap. J.}, {\bf 545}, L5
\medskip

\noindent Hoyle,F. and Narlikar, J.V. 1966, {\it Proc. Roy. Soc. A.}, {\bf 290}, 143
\medskip

\noindent Hoyle,F., Burbidge, G. and Narlikar, J.V. 1993, {\it Ap. J}, {\bf 410}, 437
\medskip

\noindent Hoyle,F., Burbidge, G. and Narlikar, J.V. 1994$a$, {\it MNRAS}, {\bf 267}, 1007
\medskip

\noindent Hoyle, F., Burbidge, G. and Narlikar, J.V. 1994$b$, {\it Astron. \& Ap.}, {\bf 289}, 729
\medskip

\noindent Hoyle,F., Burbidge, G. and Narlikar, J.V. 1995, {\it Proc. Roy. Soc. A.}, {\bf 448}, 191
\medskip

\noindent Hoyle,F., Burbidge, G. and Narlikar, J.V. 2000, {\it A Different Approach to Cosmology}, Cambridge : Cambridge University Press
\medskip

\noindent Nabarrow, F.R.N. and Jackson, P.J. 1958, in {\it Growth and Perfection in Crystals}, Eds R.H. Doremus, P.W. Roberts and D. Turnbull, Wiley, 65 
\medskip

\noindent Narlikar, J.V., Wickramasinghe, N.C. and Edmunds, M.G. 1975, in {\it Far Infrared Astronomy}, Ed. M.Rowan-Robinson, Pergaman, 131
\medskip

\noindent Narlikar,J.V., Wickramasinghe, N.C., Sachs, R. and Hoyle, F. 1997, {\it I.J. Mod. Phys}. {\bf D6}, 125
\medskip

\noindent Narlikar, J.V., Vishwakarma, R.G. and Burbidge, G. 2002,
{\it P.A.S.P.}, {\bf 114}, 1092, (astro-ph/0205064).
\medskip

\noindent  Nayeri, A., Engineer, S., Narlikar, J.V. and Hoyle, F. 1999, {\it Ap. J.}, {\bf 525}, 10
\medskip

\noindent Peebles, P. J. E.  1980, {\it The Large-Scale Structure of the Universe}, Princeton.
\medskip

\noindent Peebles, P. J. E. and Hauser, M. G. 1974, 
{\it Astrophysical Journal Supplement}, {\bf 28}, 19.
\medskip

\noindent Perlmutter, S., et al. 1999, {\it Ap. J.}, {\bf 517}, 565
\medskip

\noindent Podariu, S., Souradeep, T.,   Gott III, J. R., 
Ratra, B.  and Vogeley, M. S. 2001, {\it Astrophysical Journal}, {\bf
  559}, 9-22.
\medskip

\noindent Riess, A., et al. 1998, {\it A. J.}, {\bf 116}, 1009
\medskip

\noindent Sachs, R., Narlikar, J.V. and Hoyle, F. 1996, {\it Astronomy and Astrophysics}, {\bf 313},
703
\medskip

\noindent Smoot, G.  et al. 1992, {\it Astrophysical Journal}, {\bf 396}, L1.
\medskip

\noindent Sugiyama, N. 1995, {\it Astrophysical Journal Supplement}, {\bf 100}, 281. 
\medskip

\noindent Vishwakarma, R.G. 2002, {\it MNRAS}, {\bf 331}, 776.
\medskip

\noindent Weinberg, S. 1972, {\it Gravitation and Cosmology}, John Wiley, p.525.
\medskip


\end{document}